\documentstyle[preprint,aps]{revtex}
\begin{document}
\draft
\preprint{}
\title{ Vacuum structure and effective potential at finite temperature:\\
a variational approach}
\author{Amruta Mishra\footnote{
Present Address: 
Institut fur Theoretische Physik, Universitat Frankfurt,
Robert Mayer Str.10, D-60054, Frankfurt am Main, Germany,
e-mail: mishra@th.physik.uni-frankfurt.de} and
Hiranmaya Mishra\footnote{
Present Address:
Fakultat fur Physik, Universitat Bielefeld,
Universitat Strasse 25,
D-33615, Bielefeld, Germany,
e-mail: hmishra@physik.uni-bielefeld.de}}
\address{Mehta Research Institute of Mathematics\\ 
and Mathematical Sciences,
10, Kasturba Gandhi Marg,\\
Allahabad - 211 002,India}
\maketitle
\begin{abstract}
We compute the effective potential for $\phi^4$
theory with a squeezed coherent state type of construct for the ground state. 
The method essentially consists in optimising the basis at zero and
finite temperatures. The gap equation becomes identical to resumming 
the infinite series of daisy and super daisy graphs while the effective
potential includes multiloop effects and agrees with that obtained through 
composite operator formalism at finite temperature.
\end{abstract}
\narrowtext

\section{INTRODUCTION}
There has been a considerable interest in the ground state structure of
interacting quantum field theories since a long time. 
Variational methods like gaussian
effective potential (GEP) method appear to be a powerful technique to
study the same \cite{stev}. The equivalence between GEP and Bogoliubov
transformations for $\phi^4$ theory in 1+1 dimensions was demonstarted
earlier throgh an expilcit construction of the nonperturbative vacuum
as a squezeed coherent state\cite{gep} and was applied to study the vacuum
structure of O(N) symmetric Thirring model.
We shall test here such a nonperturbative variational method to obtain the
effective potential and the gap equation in $\phi^4$ theory in 3+1
dimensions. These have
been extensively studied by many authors at zero \cite{coleman} as well as
at finite temperatures \cite{dolan,kapusta,rudaz} by summing over daisy and
superdaisy diagrams. Leading and subleading contributions from multiloop
diagrams were also included to obtain the effective potential at finite
temperatures \cite{parwani,pisarski,samir,carring,zwirn}, where, it became
often necessary to drop various finite and divergent contributions. This
was partially circumvented in \cite{pi2,pi} while developing a self-consistent
loop expansion of the effective potential at finite temperatures.
Here the authors used composite operators \cite{cjt} and the renormalisation
prescriptions of Coleman {\it et al} \cite{politzer}. These extensive
discussions make the $\phi^4$ theory a good testing ground for examining 
a variational approximation scheme \cite{qcdt} developed essentally
for quantum chromodynamics (QCD).

The conventional calculations as stated above basically consist 
of summing over perturbative diagrams, and then considering leading
and subleading contributions arising from multiloop approximations. 
In contrast, the approximation scheme here shall use squeezed coherent 
state type construction \cite{qcdt} for the ground state which
amounts to an explicit vacuum realignment. The input here is equal
time quantum algebra with a variational ansatz for the vacuum structure,
and has no reference to perturbative expansion or Feynman diagrams.
We had earlier seen that this correctly yields the results
of Gross-Neveu model \cite{gn} as obtained by summing an infinite series of
one loop diagrams. We had also seen that it reproduces \cite{njl}
the gap equation of Nambu-Jona Lasinio model {\it without} using
Schwinger Dyson equations. We shall here apply the method to $\phi^4$ 
theory at zero and finite temperatures and show that it also yields the
self-consistent multiloop effects obtained earlier \cite{pi}. The
$\phi^4$ model has extensive applications for early universe and cosmology
\cite{linde}, and thus an understanding of the same with alternative physical
pictures is desirable.

The plan of the paper is as follows. In section II we compute the 
effective potential at zero temperature using squeezed coherent states. 
In section III we compute this at finite temperature 
using \cite{tfd,hen}
thermofield dynamics (TFD) and minimise the free energy density 
\cite{qcdt}. We shall see here that the gap equation is identical
to the one obtained by resumming the daisy and superdaisy graphs with
appropriate renormalisation \cite{dolan,pi} and the effective potential 
includes multiloop effects as through self-consistent approach with 
composite operators \cite{pi}.
In section IV we summarise the results and compare the same with
earlier calculations.

\section{Effective potential at zero temperature}
We shall compute here the effective potential for $\phi^4$ theory.
  The Lagrangian is given as

\begin{equation}
{\cal L} = {1\over 2}{\partial_{\mu}\phi\partial^{\mu}\phi}
-{1\over {2}} m^2 {\phi} ^{2}-
\lambda \phi ^{4},
\label{gep1}
\end{equation}
where, $m$ and $\lambda$ are the bare (unrenormalised) mass
and coupling constant respectively.
The fields satisfy the equal time quantum algebra
\begin{equation}
[\phi (\vec x), {\dot \phi} (\vec y)]=i \delta (\vec x-\vec y).
\label{gep2}
\end{equation}
\noindent We may expand the field operators in terms of creation and 
annihilation operators at time t=0 as
\begin{mathletters}
\begin{equation}
\phi (\vec x,0)={1\over{(2 \pi)^{3/2}} }\int{{d\vec k\over{\sqrt
{2 \omega (\vec k)}}}\left(a(\vec k)+
a(-\vec k)^{\dagger}\right)e^{i\vec k\cdot\vec x}},
\end{equation}
\begin{equation}
\dot \phi (\vec x,0)={1\over{(2 \pi)^{3/2}} }\times i
\int {d\vec k\sqrt {\omega(\vec k)\over 2}
\left(-a(\vec k)+
a(-\vec k)^{\dagger}\right)e^{i\vec k\cdot\vec x}}.
\end{equation}
\end{mathletters}
\noindent In the above, $\omega (\vec k)$ is an arbitrary function which 
for free fields is given by $\omega (\vec k)=\sqrt{\vec k^2+m^2}$ and the 
perturbative vacuum is defined corresponding to this basis through 
$a\mid vac>=0$. Further the expansions (3) and the quantum
algebra (\ref{gep2}) yield the commutation relation for the operators
$a$'s as
\begin{equation}
$$\bigl[a(\vec k), {a(\vec k')^\dagger}\bigr ]\,=\, \delta (\vec k-\vec k').
\label{gepq}
\end{equation}

Earlier a redefinition of $\mid vac>$ to $\mid vac'>$
through coherent \cite{5spm87,coherent}
or squeezed coherent states \cite{qcdt,gn}
was seen as equivalent to a Bogoliubov transformation.  We shall
adopt a similar procedure here to calculate the effective potential.
We shall take the ansatz for the trial ground 
state as \cite{gep}

\begin{equation}
\mid vac'>=U\mid vac>\equiv U_{II}U_{I}\mid vac>, 
\label{gep5}
\end{equation}
\noindent $U_{i}=\exp({B_i}^{\dagger}\,-\,B_i),\,(i=I,II)$. Explicitly
$B_{i}$'s are given as
\begin{mathletters}
\begin{equation}
{B_I}^\dagger=\int {d\vec k \sqrt{\omega (\vec k)\over 2} f(\vec k)
a(\vec k)^\dagger},
\end{equation}
and
\begin{equation}
{B_{II}}^ {\dagger}={1\over 2}\int{d\vec k g(\vec k){a'(\vec k)^\dagger}
{a'(-\vec k)^\dagger}}. 
\end{equation}
\end{mathletters}
In the above, $a'(\vec k)=U_I a(\vec k) U_I^{-1}=a(\vec k)-
\sqrt{\frac{\omega (\vec k)}{2}}f(\vec k)$ corresponds to a
shifted field operator associated with the coherent state \cite{higgs} and 
satisfies the same quantum algebra as given in equation (\ref {gepq}).
\noindent Thus in this construct for the ground state we have two
functions $f(\vec k)$ and $g(\vec k)$ which will get determined through
 minimisation of energy density. Further, since $\mid vac'>$ contains arbitrary
number of $a'^{\dagger}$ quanta, $a'\mid vac'>\,\not=\,0$. However, we
can define the basis $b(\vec k)$, $b(\vec  k)^{\dagger}$ corresponding
to $\mid vac'>$ through the Bogoliubov transformation as
\begin{eqnarray}
\left(
\begin{array}{c}
 b(\vec k)\\ b(-\vec k)^{\dagger}
\end{array}
\right)
& = & U _{II}
\left(
\begin{array}{c}
a'(\vec k)\\ a'(-\vec k)^{\dagger}
\end{array}
\right)
U_{II}^{-1}=
\left(
\begin{array}{cc}
\cosh\! g
& -\sinh\! g \\ -\sinh\!g & \cosh\!g 
\end{array}
\right)
\left(
\begin{array}{c}
a'(\vec k)\\ a'(-\vec k)^{\dagger}
\end{array}
\right).
\label{gep7}
\end{eqnarray}
\noindent It is easy to check that $b(\vec k)\mid vac'>=0 $. 
Further, to preserve 
translational invariance $f(\vec k)$ has to be  proportional to $\delta
(\vec k)$ and  we shall take $f(\vec k)=\phi _{0}  
\delta (\vec k)  \times (2\pi)^{3/2}$. $\phi_0$ will correspond to
classical field of the conventional approach \cite{higgs}.
We next calculate the expectation value of the Hamiltonian density
given as
\begin{equation}
{\cal T}^{00}=\bigl[{1 \over 2}\{(\dot {\phi })^{2}+(\vec \bigtriangledown 
\phi )^{2}+ m^2 \phi ^2 \}+\lambda {\phi }^{4}\bigr].
\label{gep8}
\end{equation}
Using the transformations (\ref{gep7}) it is easy to evaluate that
\begin{mathletters}
\begin{equation}
<vac'\mid \phi \mid vac'>=\phi _{0},
\label{gep9a}
\end{equation}
but,
\begin{equation}
<vac'\mid {{\phi(\vec z)}^2} \mid vac'>={\phi _{0}}^{2}+I,
\label{gep9b}
\end{equation}
\noindent where $I$ is the integral given as, with $k=|\vec k|$,
\begin{equation}
I={1 \over (2 \pi)^3}\int{{d\vec k \over {2\;\omega (\vec k)}}
(\cosh \!2g +\sinh\!2g)}.
\label{gep9c}
\end{equation}
\end{mathletters}
\noindent Using equations (\ref{gep8}) and (9)
the energy density of the
trial state becomes \cite{gep}
\begin{eqnarray}
<vac'\mid{\cal T}^{00}\mid vac'> &=&
\frac{1}{2}{1 \over {(2 \pi)^3}} \int{d \vec k
\over 2\omega (k)}k^{2}(\sinh\!2g +\cosh\!2g)
\nonumber\\
&+&\frac{1}{2}{1 \over {(2 \pi)^3}} \int d \vec k
\frac{\omega (k)}{2}(\cosh\!2g -\sinh\!2g)
\nonumber\\
&+& \frac{1}{2}m^2I+6\lambda\phi_0^2 I+3\lambda I^2
+\frac{1}{2}m^2\phi_0^2 +\lambda \phi_0^4,
\label{en}
\end{eqnarray}
where $I$ is as given in equation (\ref{gep9c}).
\noindent Extremising the above energy density with respect to
the function $g(\vec k)$ now yields for extremised $g(\vec k)$ as
\begin{equation}
\tanh\!{2 g(\vec k)}=-\,{{6 \lambda I+6 \lambda {\phi _0}^2}\over {
{\omega (k)}^{2}+6 \lambda I+6 \lambda {\phi _{0}}^{2}}}.
\label{gk}
\end{equation}
\noindent Substituting this value of $g(\vec k)$ in the expression
for energy density yields the effective potential as
\begin{equation}
\epsilon_0 =V(\phi_0)=
{1 \over {2}}m^2{\phi _0}^2\,+\lambda {\phi _0}^4
+\frac{1}{2}\frac{1}{(2\pi)^3}\int d\vec k (k^2+M^2)^{1/2}
-3 \lambda I^2
\label{pot}
\end{equation}
\noindent where 
\begin{equation}
M^2=m^2+3\lambda I +12\lambda \phi_0^2
\label{m2}
\end{equation}
with
\begin{equation}
I=\frac{1}{(2\pi)^3}\int\frac{d\vec k}{2} 
\frac{1}{(\vec k^2+M^2)^{1/2}}
\label{I}
\end{equation}
obtained from equation (\ref{gep9c}) after substituting for the condensate 
function $g(\vec k)$ as in equation (\ref{gk}).
The integrals in the equations (\ref{pot}) and (\ref{I})
are divergent, and require renormalisation. We use the
renormalisation prescription as in ref. \cite{politzer}
and thus obtain the renormalised mass $m_R$ and coupling
$\lambda_R$ through
\begin{mathletters}
\begin{equation}
\frac{m_R^2}{\lambda_R}=
\frac{m^2}{\lambda}+12I_1,
\label{mr}
\end{equation}
\begin{equation}
\frac{1}{\lambda_R}=
\frac{1}{\lambda}+12I_2(\mu),
\label{lr}
\end{equation}
\end{mathletters}
where $I_1$ and $I_2$ are divergent integrals given as, 
\begin{mathletters}
\begin{equation}
I_1=\frac{1}{(2\pi)^3}\int \frac{d \vec k}{2k},
\end{equation}
\begin{equation}
I_2(\mu)=\frac{1}{\mu^2}\int \frac{d \vec k}{(2\pi)^3}\Big (\frac{1}{2k}
-\frac{1}{2\sqrt{k^2+\mu^2}}\Big)
\end{equation}
\end{mathletters}
with $\mu$ as the renormalisation scale. Using equations (\ref{mr}) and 
(\ref{lr}) in equation (\ref{m2}), we have
the gap equation for $M^2$ in terms of the renormalised parameters
as
\begin{equation}
M^2=m_R^2+12\lambda_R\phi_0^2+12\lambda_R I_f(M),
\label{mm2}
\end{equation}
where, 
\begin{equation}
I_f(M)=\frac{M^2}{16\pi^2}ln \Big(\frac{M^2}{\mu^2} \Big).
\end{equation}
Using the above equations  we simplify equation (\ref{pot}) to obtain the
effective potential in terms of $\phi_0$ as
\begin{eqnarray}
V(\phi_0)&=&3\lambda_R\Big(\phi_0^2+\frac{m_R^2}{12\lambda_R}\Big)^2
\nonumber \\
&+&\frac {M^4}{64\pi^2}ln\Big(\frac{M^2}{\mu^2}-\frac{1}{2}\Big)
-3\lambda_R I_f^2-2\lambda\phi_0^4.
\label{vph}
\end{eqnarray}
This potential is identical to that as obtained through composite operator
formalism as in Ref. \cite{pi2} and is derived through a purely
nonperturbative approach through vacuum realignment.
We now consider the generalisation
of the above to finite temperatures.
\section{Effective potential at finite temperature}
We shall calculate the effective potential
at finite temperature for $\phi^4$ theory using thermofield dynamics
\cite{tfd}. Here the statistical average of an
operator is written as an expectation value with respect to a
``thermal vacuum" constructed from operators defined on 
an extended Hilbert space \cite{tfd}.
The ``thermal vacuum" is obtained from the zero temperature
ground state through a ``thermal" Bogoliubov transformation.
Thus at finite temperatures, $|vac'>$ of equation (5),
with $\beta$ as inverse of temperature, goes over to
\begin{equation}
|vac',\beta>=U(\beta)|vac'>,
\label{vact}
\end{equation}
where \cite{tfd}
\begin{equation}
U(\beta)=\exp(B(\beta)^\dagger-B(\beta)),
\end{equation}
with
\begin{equation}
B(\beta)^\dagger=\int d\vec k \theta(\vec k,\beta)b(\vec k)^\dagger
{\tilde b}(-\vec k)^\dagger.
\end{equation}
In the above, $\tilde b$ is the annihilation operator for {\it thermal}
modes associated with the doubling of the Hilbert space \cite{tfd}.
The function $\theta(\vec k, \beta)$ for the
corresponding Bogoliubov transformation
is related to the distribution function and shall be obtained by 
minimising free energy density. We first note that the
temperature dependant annihilation operators
$b(\vec k,\beta)$ and $\tilde b(\vec k,\beta)$ corresponding to the
thermal vacuum  are given as
\begin{eqnarray}
\left(
\begin{array}{c}
 b(\vec k,\beta)\\ \tilde b(-\vec k,\beta)^{\dagger}
\end{array}
\right)
& = & U (\beta)
\left(
\begin{array}{c}
b(\vec k)\\ \tilde b(-\vec k)^{\dagger}
\end{array}
\right)
U(\beta)^{-1}=
\left(
\begin{array}{cc}
\cosh\! \theta
& -\sinh\! \theta \\ -\sinh\!\theta & \cosh\!\theta 
\end{array}
\right)
\left(
\begin{array}{c}
b(\vec k)\\ \tilde b(-\vec k)^{\dagger}
\end{array}
\right).
\label{bogt}
\end{eqnarray}
Using the transformations given by (\ref{gep7}) and (\ref{bogt}),
we can calculate the expectation values of the operators in the thermal
vacuum. For example, the expectation value of $\phi^2$ of equation
(\ref{gep9b}) in thermal vacuum becomes

\begin{equation}
<vac',\beta\mid {{\phi(\vec z)}^2} \mid vac',\beta>=
{\phi _{0}}^{2}+I(\beta),
\label{gep9bt}
\end{equation}
\noindent where $I(\beta)$ is the integral
\begin{equation}
I(\beta)={1 \over (2 \pi)^3}\int{{d\vec k \over {2\;\omega (\vec k)}}
(\cosh \!2g +\sinh\!2g)}\cosh\! 2\theta(\vec k,\beta),
\label{gep9ct}
\end{equation}
parallel to equation (\ref{gep9c}) at zero temperature. 
The energy density then becomes
the generalisation of equation (\ref{en}),
given as
\begin{eqnarray}
\epsilon\equiv <vac',\beta \mid{\cal T}^{00}\mid vac',\beta > &=&
\frac{1}{2}{1 \over {(2 \pi)^3}} \int{d \vec k
\over 2\omega (k)}k^{2}(\sinh\!2g +\cosh\!2g)\cosh\!2\theta
\nonumber\\
&+&\frac{1}{2}{1 \over {(2 \pi)^3}} \int d \vec k
\frac{\omega (k)}{2}(\cosh\!2g -\sinh\!2g)\cosh\!2\theta
\nonumber\\
&+& \frac{1}{2}m^2I(\beta)+6\lambda\phi_0^2 I(\beta)+3\lambda I(\beta)^2
+\frac{1}{2}m^2\phi_0^2 +\lambda \phi_0^4.
\label{ent}
\end{eqnarray}
At finite temperature the relevant quantity to be
extremised is free energy density given as
\begin{equation}
{\cal F}(\phi_0,g,\theta)=\epsilon-\frac {1}{\beta} S,
\end{equation}
with the entropy density $S$ as \cite{tfd}
\begin{equation}
S=\frac {1}{(2\pi)^3} \int d \vec k \bigl(
\cosh^2 \! \theta ln(\cosh^2\!\theta)-
\sinh^2 \! \theta ln(\sinh^2\!\theta)\bigr)
\label{entr}
\end{equation}
Minimising ${\cal F}
(\phi_0,g,\theta)$ with respect to $g(\vec k)$ we obtain that
\begin{equation}
\tanh\!{2 g(\vec k,\beta)}=-\,{{6 \lambda I(\beta)+6 \lambda {\phi _0}^2}\over {
{\omega (k)}^{2}+6 \lambda I(\beta)+6 \lambda {\phi _{0}}^{2}}}.
\label{gkt}
\end{equation}
Further, extremising the free energy density with respect to
$\theta(\vec k,\beta)$ we obtain that 
\begin{equation}
\sinh ^2 \! \theta(\vec k,\beta)
=\frac{1}{e^{\beta \omega '(\vec k,\beta)}-1}
\label{sth}
\end{equation}
where, 
\begin{equation}
\omega ' (\vec k,\beta)=({\vec k}^2+M(\beta)^2)^{1/2}
\end{equation}
with
\begin{equation}
M(\beta)^2=m^2+3\lambda I(\beta) +12\lambda \phi_0^2.
\label{m2t}
\end{equation}
The above equation is the finite temperature generalisation of (\ref{m2})
and corresponds to the gap equation after summing over daisy and superdaisy
graphs \cite{dolan,pi}. Substituting the optimised 
expressions for the functions $\theta(\vec k,\beta)$ and
$g(\vec k,\beta)$, 
the free energy density or the effective potential at finite
temperature now becomes
\begin{equation}
V(\phi_0,\beta)\equiv {\cal F}=V_0+V_I+V_{II}
\label{pott}
\end{equation}
with,
\begin{mathletters}
\begin{equation}
V_0={1 \over 2}m^2{\phi _0}^2\,+\lambda {\phi _0}^4,
\end{equation}
\begin{equation}
V_I=\frac {1}{2}\frac {1}{(2\pi)^3}\int d\vec k (k^2+M(\beta)^2)^{1/2}
\cosh \! 2\theta
-\frac {1}{\beta}S
\end{equation}
and,
\begin{equation}
V_{II}=-3 \lambda I(\beta)^2,
\end{equation}
\end{mathletters}
where
\begin{equation}
I(\beta)=\frac{1}{(2\pi)^3}\int\frac{d\vec k}{2} 
\frac{\cosh \! 2\theta}{(\vec k^2+M(\beta)^2)^{1/2}}.
\label{It}
\end{equation}
We may note that in equation (\ref{pott}) $V_0$ is the tree level potential,
and, $V_I$ and $V_{II}$ can be identified with the one loop and two loop
contributions of Ref. \cite{pi} {\em after} summing over the discrete
frequencies corresponding to imaginary time formulation of finite temperature
field theory. The gap equation and the effective potential here 
have been obtained by minimising free energy with a nontrivial vacuum 
structure as in equation (\ref{vact}).

We note that the finite temperature contributions do not bring in any 
fresh divergences apart from those encountered at zero temperature.
We thus use the renormalisation conditions as in equation (15). 
In that case, the divergences from $V_0$ and $V_{II}$ combine to cancel 
the same arising from $V_I$ when equation (\ref{sth}) for the distribution
function is used \cite{pi}. The {\it renormalised} effective potential then 
becomes
\begin{equation}
V(\phi_0,\beta)
= V_0+V_I,+V_{II}
\label{vpht}
\end{equation}
where,
\begin{equation}
V_0+V_{II}=
3\lambda_R\Big(\phi_0^2+\frac{m_R^2}{12\lambda_R}\Big)^2
- 3\lambda_R I_f(\beta)^2-2\lambda\phi_0^4
\end{equation}
and,
\begin{equation}
V_I=
\frac {M(\beta)^4}{64\pi^2}\Big(ln(\frac{M(\beta)^2}{\mu^2})
-\frac{1}{2}\Big)
+\frac{1}{\beta} \frac{1}{(2\pi)^3}\int d \vec k ln \left (1-\exp(-\beta\omega')
\right ).
\end{equation}
In the above $M(\beta)$ satisfies the renormalised gap equation
\begin{equation}
M(\beta)^2=m_R^2+12\lambda_R\phi_0^2+12\lambda_R I_f(M(\beta)),
\label{mm2t}
\end{equation}
where, 
\begin{equation}
I_f(M(\beta))=\frac{M(\beta)^2}{16\pi^2}
ln \Big(\frac{M(\beta)^2}{\mu^2}\Big)
+\int \frac{d\vec k}{(2\pi)^3} 
\frac{\sinh ^2 \! \theta(\vec k,\beta)}{({\vec k}^2+M(\beta)^2)^{1/2}}.
\end{equation}
We note that the above expressions for the effective potential as well
as the gap equation (39) are identical respectively to equations (3.19),
(3.20) and (3.16) of Ref. \cite{pi}. Thus the realignment of the
ground state with condensates naturally generates the nonperturbative
features beyond one loop as demonstrated above.

\section{Discussions}
We first note that equation (\ref{vpht}) is the same as the earlier
equations of Ref. \cite{pi}, and hence the same conclusions
regarding high temperature limits as well as discussions on the
nature of the phase transitions continue to hold good with the present
picture of phase transition through explicit vacuum realignment.
In particular we may also note that the functions arising from
composite operators of Ref. \cite{pi} on summing over discrete frequencies
correspond to $I(\beta)$ of equation (\ref{It}) arising from the variation 
over the condensate function. The thermal distribution functions
{\it including interactions} as 
in equation (\ref{sth}) has been derived through an
{\it extremisation} of the free energy density.
Coherent states \cite{hen} as well as squeezed states 
for dissipative systems \cite{vit1} along with
thermofield dynamics have been dealt with for quantum mechanical
problems. Here we have applied the techniques to quantum
field theory to include nonperturbative effects.
Vacuum effects have been known to be relevant for low energy
nonperturbative physics \cite{svz,suryak}. We have also seen that
a nontrivial vacuum structure with condensates is not only conceptually
different, but can have phenomenological implications 
\cite{qcdt,njl,amspm,vit2}. 
Thus the present work with an explicit structure
for thermal vacuum may become relevant for phenomenology in cosmology.

It is interesting to note that the present variational ansatz
with squeezed vacuum structure leads to daisy- super daisy
resummed self consistent two loop effective potential
as obtained in Ref.\cite{pi}. The reason for  such a result
lies on the fact that the $\phi^4$ interaction leads to a
functional for vacuum energy which is effectively quadratic 
and we could solve for the ansatz functions explicitly.
With a more complicated structure for the ansatz state
involving nonlinear canonical transformations one can improve
upon the effective potential as calculated here \cite{nlct}.
The inputs have been 
(i) equal time algebra of interacting fields, (ii) an ansatz for
vacuum realignment, and, (iii) use of renormalisation prescriptions of
Ref. \cite{politzer}. We did not have to go through the perturbative
route anywhere, and no summation of diagrams was needed. The results thus 
obtained went beyond the one loop approximation. However,
since variational methods are ansatz dependant, it is nice to see
that the present picture more easily reproduces the
final features of a well studied problem, and adds conceptual
ingredients that were absent in the earlier calculations which
could be relevant for phenomenology. The analysis thus adds to the confidence
with which the present method can be applied, as well as gives fresh
understanding of summation of infinite diagrams of perturbation series.

\acknowledgements The authors wish to acknowledge discussions with
S.P. Misra. They also acknowledge to the Council of Scientific and
Industrial Research, Government of India for the research associateships
9/679(3)/95-EMR-I and 7/679(3)/95-EMR-I.

\end{document}